\documentclass[aps,prl,twocolumn,superscriptaddress]{revtex4}%
\usepackage{amsfonts}
\usepackage{amsmath}
\usepackage{amssymb}
\usepackage{graphicx}
\usepackage{xcolor}
\usepackage{epstopdf}
\usepackage{bm}%
\begin{document}
\title{Stealth Majorana Zero Mode in a Trilayer Heterostructure MnTe/Bi$_{2}$Te$_{3}$/Fe(Te,Se)}
\author{Rui Song}
\affiliation{Science and Technology on Surface Physics and Chemistry Laboratory, Mianyang,
Sichuan 621908, China}
\affiliation{Anhui Key Laboratory of Condensed Matter Physics at Extreme Conditions, High
Magnetic Field Laboratory, HFIPS, Anhui, Chinese Academy of Sciences, and
University of Science and Technology of China, Hefei 230031, China}
\author{Ning Hao}
\email{haon@hmfl.ac.cn}
\affiliation{Anhui Key Laboratory of Condensed Matter Physics at Extreme Conditions, High
Magnetic Field Laboratory, HFIPS, Anhui, Chinese Academy of Sciences, and
University of Science and Technology of China, Hefei 230031, China}

\begin{abstract}
Recent experiment reported the robust zero-energy states with strange
properties in a trilayer heterostructure MnTe/Bi$_{2}$Te$_{3}$/Fe(Te,Se).
Here, we give comprehensive understandings about the magnetic and electronic
properties of the heterostructure, and propose ferromagnetic Mn-Bi antisite
defects are generated in the topmost sublayer of Bi$_{2}$Te$_{3}$ and hidden
below the MnTe layer. We further reveal the defect can induce two types of
quasiparticles. One is Yu-Shiba-Rusinov state from defect itself, and another
is Majorana zero mode from the superconducting phase domain wall induced by
the defect. The two types of quasiparticles have very different response to
magnetic field, temperature etc. The coexistence and mutual cooperation of
both can explain experimental observations. Furthermore, we propose more
simple heterostructure with superiority to generate and finely modulate
Majorana zero modes.

\end{abstract}
\maketitle

Since the discovery of high-temperature superconductivity in FeSe/SrTiO$_{3}%
$\cite{FeSe-1}, the heterostructures involving Fe(Te,Se) have attracted
intense interests to explore the new interfacial physical phenomena, such as
spin density wave, interfacial phonon mode, topological physics, gating
effect, competition between superconductivity and magnetism
etc\cite{FeSe-2,FeSe-3,FeSe-4,FeSe-5,FeSe-6,FeSe-7,FeSe-8,FeSe-9}. Recently,
an unexpected robust zero-energy state was experimentally observed in a
trilayer heterostructure of MnTe/Bi$_{2}$Te$_{3}$/Fe(Te,Se)\cite{FeSe-10}, and
the state shows some confusing properties, such as invisuality from surface
topography but visuality from zero-bias dI/dV map, the robustness against
external magnetic field, strange temperature evolution relation etc.
Considering both Bi$_{2}$Te$_{3}$ and Fe(Te,Se) are two kinds of well-known
topological materials, it is natural to connect the robustness of zero-energy
state with the nontrivial topology of the system. However, the complexity of
the trilayer heterostructures blurs and obscures such connection. Thus, the
following urgent issues arise: is the robust zero-energy state a Majorana zero
mode (MZM)? how is it generated? how to understand its strange properties?
Clarification of these issues could establish such trilayer heterostructure as
a new platform to study physics of MZMs, or supply an inspired strategy to
construct more pragmatic system to realize MZMs.

In this work, we address the aforementioned issues through comprehensive
understanding the magnetic and electronic properties of the trilayer
heterostructures with density functional theory (DFT) calculations and
extracting the underlying physics of zero-energy state with a simplified
theoretical model. Our DFT calculations suggest single unit cell (UC) MnTe
could probably generate weak antiferromagnetic (AFM) fluctuation rather than
long-range AFM order. The consequent band structures are consistent with the
experimental results from quasiparticle interference (QPI) measurement. Then,
we propose there exists at least two different kinds of point defects in the
trilayer heterostructure. One is from the vacancy of Te atom on the top
surface of 1-UC MnTe/Bi$_{2}$Te$_{3}$/Fe(Te,Se). Another is the Mn-Bi antisite
defect in form of Mn atom substituting Bi atom in top layer of Bi$_{2}$%
Te$_{3}$. The Mn-Bi antisite defect will generate a magnetic moment to couple
the spin of the surface electrons of Bi$_{2}$Te$_{3}$ with ferromagnetic type.
Such Mn-Bi antisite defect associated with ferromagnetic coupling not only
induces the trivial Yu-Shiba-Rusinov (YSR) state but also drives a local
quantum phase transition (QPT) of the Fe(Te,Se)-proximity-effect-enabled
surface superconductivity of Bi$_{2}$Te$_{3}$. Namely, the superconducting
order parameter of Bi$_{2}$Te$_{3}$ around the defect will acquire a
sign-change beyond the QPT. Then, a degenerate pair of MZMs protected by an
emergent mirror symmetry is induced and trapped by the phase domain wall near
the Mn-Bi antisite defect. YSR state and MZM have very different response to
the external magnetic field, temperature etc. Coexistence and mutual
cooperation of both can explain all the experimentally observed confusing
properties of the quasi-paricles lying in the superconducting gap. For the
trilayer heterostructure with 2-UC MnTe, we find that the aforementioned
picture is violated, and only trivial results can be obtained. Furthermore, we
predict unexplored properties of the zero-energy state and propose more simple
platforms to generate and modulate MZMs.\begin{figure}[pt]
\begin{center}
\includegraphics[width=1.0\columnwidth]{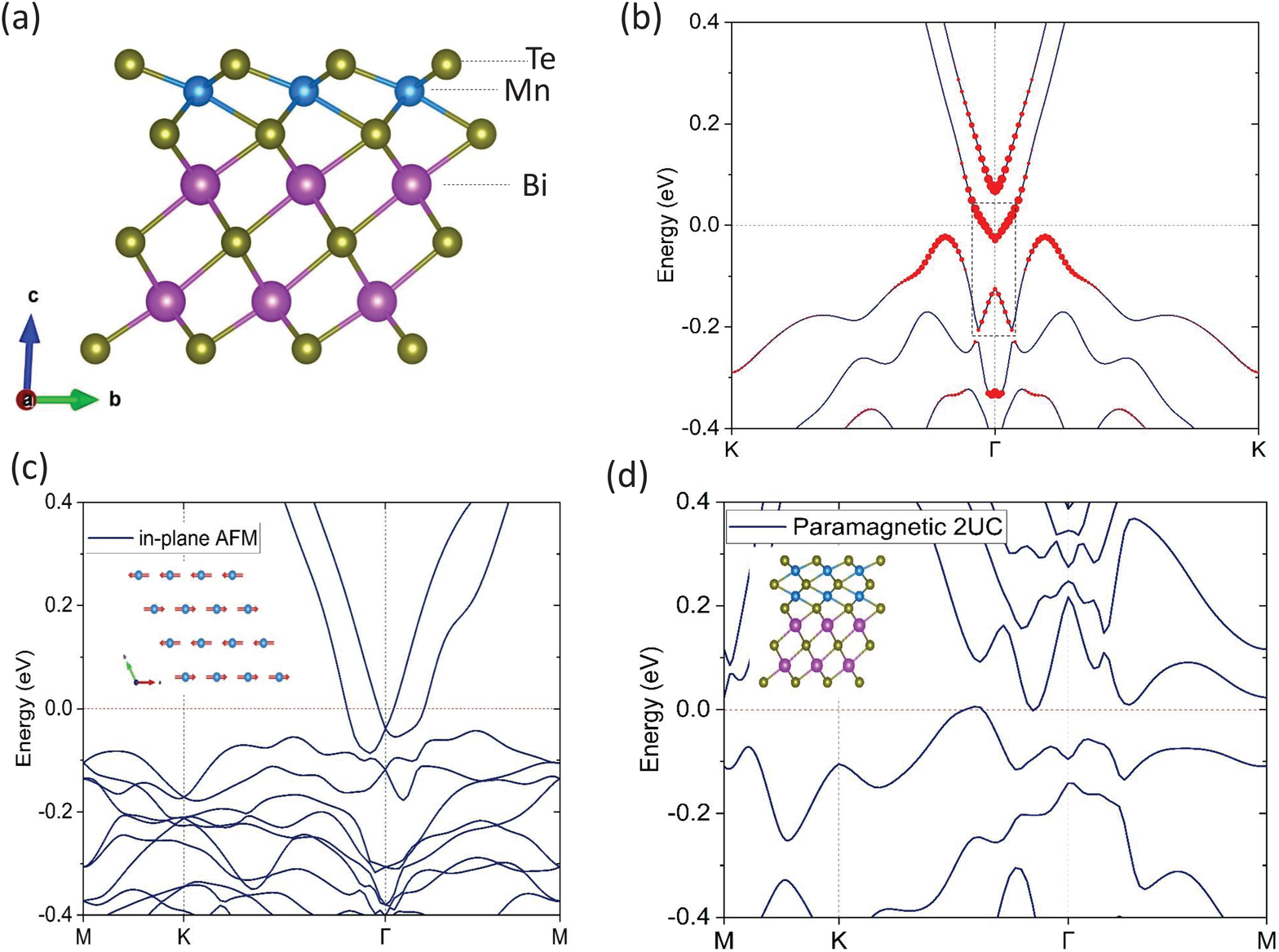}
\end{center}
\caption{(a) The heterostructure of MnTe/Be$_{2}$Te$_{3}$. (b) The band
structure of heterostructure without long-range magnetic order in MnTe. The
sizes of the red dots label the weight from the topmost sublayer of Be$_{2}%
$Te$_{3}$. The bands in the rectangle regime are captured by Eq. 2. (c) The
band structure of heterostructure with in-plane AFM long-range order in MnTe.
Insert gives the pattern of AFM order. (d) The band structure of
heterostructure involving 2-UC MnTe in the absence of any long-range magnetic
order. Insert gives the configuration of the heterostructure.}%
\label{fig-band}%
\end{figure}

As verified by the experiment, the 1-UC MnTe shares the same crystal constant
with 1-UC Bi$_{2}$Te$_{3}$. In DFT calculations, we treat 1-UC Bi$_{2}$%
Te$_{3}$ as a substrate and neglect the influence of bare Fe(Te,Se) substrate,
since Fe(Te,Se) weakly modulates the band structures of Bi$_{2}$Te$_{3}$
except supplying proximity superconductivity\cite{BiTe-FeTeSe-1,BiTe-FeTeSe-2}%
. Let the MnTe layer relax freely until the force and energy criterion is
satisfied. The final structure is shown in Fig. \ref{fig-band} (a). Te-Mn-Te
sandwich structure is as same as it in $\alpha$-MnTe
crystal\cite{alpha-MnTe-1,alpha-MnTe-2}. Take magnetic order of bulk $\alpha
$-MnTe as a reference, various magnetic orders for 1-UC MnTe are considered.
Our DFT results are consistent with the calculations in Ref. \cite{FeSe-10}.
Furthermore, we calculate the band structures of 1-UC MnTe/Bi$_{2}$Te$_{3}$ by
assuming several different kinds of long-range AFM or FM orders (See
SMs\cite{SM} for details). Fig. \ref{fig-band} (c) shows that the bands
crossing Fermi level around $\Gamma$ point split into Rashba-type, which is
inconsistent with the parabolic-type extracted by QPI
measurement\cite{FeSe-10}. Thus, we propose there only exist weak AFM
fluctuation in 1-UC MnTe, even though the DFT predict 1-UC MnTe with AFM-1
order has the lowest energy. We argue such divergence is not rare in
low-dimensional systems. For example, DFT predicts checkerboard AFM order in
monolayer FeSe/SrTiO$_{3}$%
\cite{FeSe-mag-theo-1,FeSe-mag-theo-2,FeSe-mag-theo-3,FeSe-mag-theo-4}, but no
experiment observes such order
\cite{FeSe-mag-exp-1,FeSe-mag-exp-2,FeSe-mag-exp-3,FeSe-mag-exp-4,FeSe-mag-exp-5,FeSe-mag-exp-6}%
. The observation of gapless Dirac-cone-like surface state in MnBi$_{2}%
$Te$_{4}$ also hints the missing long-rang magnetic order in topmost MnTe
layer\cite{MnBiTe-1,MnBiTe-2,MnBiTe-3,MnBiTe-4}. Our calculations also find
tiny magnetic anisotropy in 1-UC MnTe, which further indicates the missing
long-rang AFM order in 1-UC MnTe according to Mermin-Wagner
theorem\cite{M-W-theo}. Furthermore, the experiment observed the enhancement
of superconductivity in 1-UC MnTe trilayer heterostructure. We propose such
enhancement comes from the gain effect of AFM fluctuation in 1-UC MnTe
accoring to spin-fluctuation-mediated superconductivity theory. Here, we do
not extend the discussion about enhancement of superconductivity and will
address it elsewhere. Once the long-range AFM order is missing, the calculated
band structure of 1-UC MnTe/Bi$_{2}$Te$_{3}$ shown in Fig. \ref{fig-band} (b)
has a simple parabolic-type band crossing Fermi level, which is consistent
with result of QPI measurement\cite{FeSe-10}. Thus, we conclude that the first
effect of 1-UC MnTe in trilayer structure is to introduce AFM fluctuation to
enhance superconductivity and the second is to change the Fermi level by
charge doping. Fig. \ref{fig-band} (d) shows only trivial band structure
survives in the case with 2-UC MnTe. We do not further discuss it.
\begin{figure}[pt]
\begin{center}
\includegraphics[width=1.0\columnwidth]{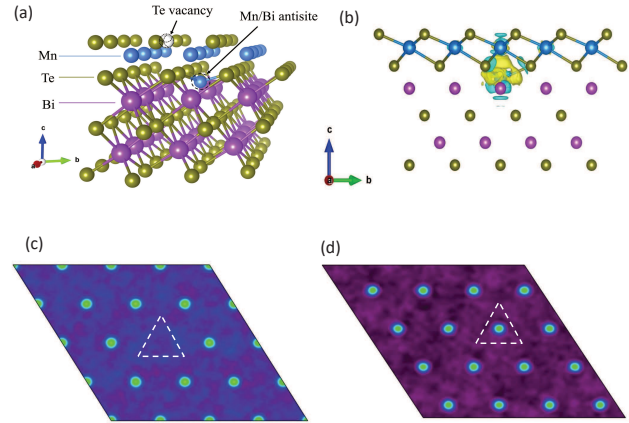}
\end{center}
\caption{(a) Two types of defects are schematically shown in the
heterostructure. One is Te vacancy in the topmost sublayer of MnTe and another
is Mn-Bi antisite defect in the topmost sublayer of Be$_{2}$Te$_{3}$. (b) The
differential charge density distribution around the Mn-Bi antisite defect.
(c)-(d) The simulated STM images on topmost surface of the heterostructure
including a Te vacancy in (c) and a Mn-Bi antisite defect in (d). }%
\label{fig-defect}%
\end{figure}

Remarkably, two types of Mn-Bi and Bi-Te antisite defects were experimentally
verified in MnBi$_{2}$Te$_{4}$\cite{Mn-impurity}. It is natural to expect
similar antisite defects could exist on the interface between 1-UC MnTe and
1-UC Bi$_{2}$Te$_{3}$ in such heterostructure. In particular, the
heterostructure is bulit with the molecular beam epitaxy (MBE) method. Fig.
\ref{fig-defect} (a) shows structures of two types of typical defects, i.e.,
the defect on the topmost sublayer 1-UC MnTe formed by Te vacancy and the
Mn-Bi antisite defect with a Mn atom substituting a Bi atom on the topmost
sublayer 1-UC Bi$_{2}$Te$_{3}$. Fig. \ref{fig-defect} (b) shows the calculated
differential charge density distribution around the Mn-Bi antisite defect. It
is obvious the charge transfer mainly happens between the bottommost sublayer
of 1-UC MnTe and the topmost layer of 1-UC Bi$_{2}$Te$_{3}$. Then, charge
density distribution is nearly not affected on the topmost surface of
heterostructure. It indicates the STM scanning topographic image is not easy
to find and fix positions of the Mn-Bi antisite defects. Fig. \ref{fig-defect}
(c)-(d) shows the DFT simulated STM image, where only Te-vacancy defect can be
observed. If the observed zero-energy states have close relationship with the
Mn-Bi antisite defects, the randomness of them can be easily understood now.
Our calculations also find that the Mn-Bi antisite defect spontaneously
generates a local magnetic moment of 4.8 $\mu_{B}$ due to the magnetism of Mn
atom. In the following, we will show that the ferromagnetic Mn-Bi antisite
defects play a key role to form the observed zero-energy states, which are
indeed the MZMs. Since Mn-Bi antisite defects in the inner layer of trilayer
heterostructure cannot be detected by the outer-layer charge distribution but
by the defect-bound superconducting quasi-particle tunnel spectrum, we call
the zero-energy states the stealth MZMs. Thus, we conclude the third effect of
1-UC MnTe in trilayer structure is to introduce ferromagnetic Mn-Bi antisite
defects in the topmost sublayer 1-UC Bi$_{2}$Te$_{3}$ in the trilayer heterostructure.

With the above prepared information, we can construct an effective model to
answer the question how to generate the robust zero-energy state. The
effective Hamiltonian can be expressed as follows,%

\begin{equation}
H_{tot}=H_{sur}+H_{d}+H_{\Delta}. \label{H-eff}%
\end{equation}
Here,%

\begin{align}
H_{sur}  &  =v_{F}(k_{x}\sigma_{y}-k_{y}\sigma_{x})-(\mu-Ak^{2})+(\Lambda
_{0}-Bk^{2})\sigma_{z}\kappa_{z},\label{H-s}\\
H_{d}  &  =\int d\mathbf{r}J(\mathbf{r})\mathbf{S}_{def}\mathbf{\cdot
\mathbf{\bm{\sigma}}},\text{ }H_{\Delta}=\Delta(\mathbf{r})\tau_{x}.
\label{H-dd}%
\end{align}
Here, $H_{sur}$ is the low-energy effective Hamiltonian to describe the
surface bands of 1-UC MnTe/Bi$_{2}$Te$_{3}$ thin film\cite{TI-model}, as
indicated by the red-dashed rectangle in Fig. \ref{fig-band} (b). Note that
the effect of structural inversion asymmetry is neglected, because the
Rashba-like splitting induced by such effect is inconsistent with the QPI
measurement\cite{FeSe-10}. $H_{d}$ describes the coupling between the
ferromagnetic Mn-Bi antisite defect and surface electron. $H_{\Delta}$ is the
proximity-effect-induced superconducting pair term from Fe(Te,Se). $v_{F}$ is
Fermi velocity. $\mu$ is chemical potential. $\Lambda_{0}$ measures gap of
Dirac cone opened by finite-thickness effect. $A$ and $B$ are coefficients of
the second-order terms. $J(\mathbf{r})=J\delta(\mathbf{r})$ describing the
ferromagnetic coupling with a constant $J$. $\mathbf{S}_{def}$ is the moment
of defect with only $z$-component. $\Delta(\mathbf{r})$ is the superconducting
order parameter from proximity effect. Note that $\Delta(\mathbf{r})$ is a
constant without the defect. $\sigma$, $\kappa$ and $\tau$ are Pauli matrix to
span spin, psudo-valley and particle-hole subspaces, respectively. $H_{tot}$
can be numerically solved in the lattice case\cite{previous-work} (See
SMs\cite{SM} for details). The influence of ferromagnetic defect can be
understood as follows. As $J(\mathbf{r}=0)$ increases from zero,
$\Delta(\mathbf{r}=0)$ at the impurity site is depressed and goes to zero at a
critical $J_{c}$, beyond which, $\Delta(\mathbf{r}=0)$ has a sign-change
\cite{RMP-1,QPT-1,QPT-2,Order-1,Order-2,Order-3}. It means there exists a QPT
at critical $J_{c}$. Meanwhile, the hole-type and electron-type branches of
YSR states\cite{YSR-1,YSR-2,YSR-3} undergo approaching-crossing-separating
processes as $J(\mathbf{r}=0)$ increases from zero. The crucial physical
quantity $J_{c}$ can be estimated by the calculations. We find $J_{c}%
\sim4.2v_{F}k_{F}$ with $v_{F}k_{F}\sim4.5$mV measuring the energy scale of
surface Dirac cone of Bi$_{2}$Te$_{3}$. Then, $J_{c}\sim19$mV, which can be
further reduced by the atomic spin-orbit coupling.

\begin{figure}[pt]
\begin{center}
\includegraphics[width=1.0\columnwidth]{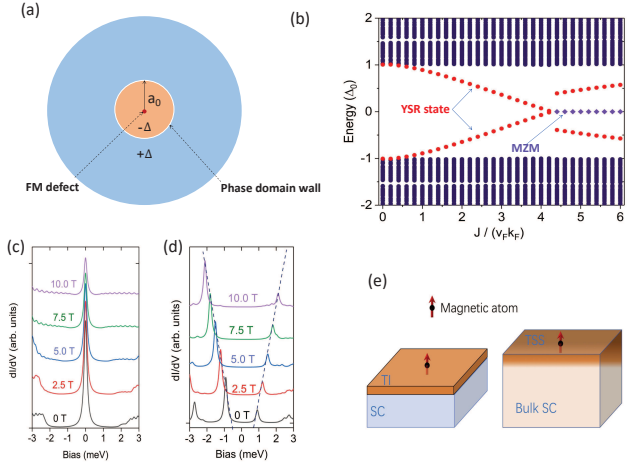}
\end{center}
\caption{(a) The 0-$\pi$ disk junction. The FM defect at center and phase
domain wall separating $\pm\Delta$ regimes are marked. $a_{0}$ is radius of
$-\Delta$ regime and approaches lattice constant. (b) The calculated overall
quasi-particle spectrum as change as $J$ under the 0-$\pi$ disk junction
geometry. The red dots label the YSR states from the ferromagnetic Mn-Bi
antisite defect in the center of the disk in (a). The purple diamonds label
the MZMs from the phase domain wall shown in (a). The trivial edge modes are
not shown (See SMs\cite{SM} for details). (c) and (d) The calculated different
response of MZM and YSR state to the external magnetic field, respectively.
(e) The proposed two kinds of platforms with magnetic defect to realized MZM.
The left one includes topological insulator and superconductor. The right one
only is a superconducting topological material.}%
\label{fig-mzm}%
\end{figure}

Besides the impurity-induced trivial YSR states, we now turn to other
quasi-particle related to the zero-energy state. Note that beyond QPT, i.e.,
$J>J_{c}$, there exists a phase domain wall for $\Delta(\mathbf{r})$ as shown
in Fig. \ref{fig-mzm} (a). Namely, $\Delta(\mathbf{r})<0$ for $r<a_{0}$ and
$\Delta(\mathbf{r})>0$ for $r>a_{0}$ with $a_{0}$ a characteristic length
close to lattice constant of Bi$_{2}$Te$_{3}$. Then, in the case of $J>J_{c}$,
$H_{tot}$ in Eq. (\ref{H-eff}) can be further reduced to the following form%

\begin{equation}
H_{r}=H_{sur}+\Delta_{s}(\mathbf{r})\tau_{x}. \label{H-r}%
\end{equation}
Here, $\Delta_{s}(\mathbf{r})=-\Delta_{1}$ for $r<a_{0}$ and $\Delta
_{s}(\mathbf{r})=\Delta_{2}$ for $r>a_{0}$ with $\Delta_{1/2}>0$ and
$0<\Delta_{1}<\Delta_{2}$. The phase of $\Delta_{1/2}$ is uniform and is
omitted. In continuum limit, the eigen-equation of $H_{r}$ is%

\begin{equation}
H_{r}(\mathbf{k}\rightarrow-i\mathbf{\nabla})\psi(r,\theta)=E\psi
(r,\theta),\label{H-eigen}%
\end{equation}
which can be solved under boundary conditions with a 0-$\pi$ disk junction
geometry as shown in Fig. \ref{fig-mzm} (a). The 0-$\pi$ disk junction can be
obtained by bending a 0-$\pi$ line junction. Such deformation enables
$\psi(r,\theta)$ to gain a phase factor $e^{i\kappa_{0}\tau_{0}\sigma
_{z}\theta/2}$\cite{previous-work1}. Thus, the wave function $\psi(r,\theta)$
must obey anti-periodic boundary condition $\psi(r,\theta+2\pi)=-\psi
(r,\theta)$\cite{previous-work,previous-work1,previous-work2}. We have
numerically proven such boundary condition\cite{previous-work}. The details to
solve Eq. (\ref{H-eigen}) are shown in SMs\cite{SM}. Without considering the
second order of $k$ terms in $H_{sur}$ of Eq. (\ref{H-s}), a pair of MZMs can
be obtained. The wave function of the first MZM takes the form $\psi
_{1}(r,\theta)=[e^{-i\theta/2}u_{\uparrow}(r),e^{i\theta/2}u_{\downarrow
}(r),e^{-i\theta/2}v_{\downarrow}(r),-e^{i\theta/2}v_{\uparrow}%
(r),0,0,0,0]^{T}$ with $u_{\sigma}(r)=-v_{\sigma}(r)$. $u_{\sigma
}(r)=a_{\sigma}J_{\mp1/2}(k_{F}r)e^{r/\xi_{1}}$ for $r<a_{0}$ and $u_{\sigma
}(r)=b_{\sigma}J_{\mp1/2}(k_{F}r)e^{-r/\xi_{2}}$ for $r>a_{0}$. $J_{\mp
1/2}(k_{F}r)$ is Bessel functions with $\mp1/2$ for spin up and down,
respectively. The Fermi wave vector is $k_{F}=\sqrt{\mu^{2}-\Lambda_{0}^{2}%
}/v_{F}$. Decay length $\xi_{1/2}=v_{F}/\Delta_{1/2}$. The wave function of
the second MZM is $\psi_{2}(r,\theta)=\mathcal{T}\psi_{1}(r,\theta)$ with
$\mathcal{T}=i\kappa_{x}\sigma_{y}\tau_{0}\mathcal{K}$ the emergent
time-reversal symmetry operator. If the second order of $k$ terms in $H_{sur}$
of Eq. (\ref{H-s}) is taken into account, the unchanged in-plane spin texture
of the surface state still gives effective topological
superconductivity\cite{helical-SC}, and a pair of MZMs is robust. In such a
case, the analytical solutions are not available, but the numerical
calculations give the clear solutions of MZMs, as shown in Fig. \ref{fig-mzm}
(b) (See SMs\cite{SM} for details).

Now, we have two types of quasi-particles in superconducting state of the 1-UC
MnTe/Bi$_{2}$Te$_{3}$/Fe(Te,Se), as shown in Fig. \ref{fig-mzm} (b). One is
trivial YSR state from the ferromagnetic Mn-Bi antisite defect itself, another
is MZM bounded by the superconducting phase domain wall induced by the defect.
These two types of quasi-particles have very different response to the
external magnetic field, temperature etc. However, the coexistence and mutual
cooperation of two types quasi-particles can explain all the observed
phenomena. The first one is to apply an external magnetic field. The energy
levels of YSR must shift through the Zeeman coupling as shown in Fig.
\ref{fig-mzm} (d). This behavior was observed by the experiment for the
non-zero-energy bound states. However, a pair of MZMs is not split under the
out-of-plane magnetic field, and the relevant forms of wave function of MZM
solutions of Eq. (\ref{H-eigen}) are unchanged except $k_{F}$ being
renormalized to$\sqrt{\mu^{2}-(\Lambda_{0}\pm\mu_{B}B_{z})^{2}}$. $\mu_{B}$
and $B_{z}$ are Bohr magneton and $z$-direction of magnetic field strength.
Our calculations show that $\mu\sim100$meV and $\Lambda_{0}\sim45$meV$\sim
375$T. The experimental applied magnetic field $<10$T has negligible effect to
the MZMs. Thus, the pair of MZMs are robust against out-of-plane magnetic
field as shown in Fig. \ref{fig-mzm} (c). Such robustness is actually
protected by an hidden mirror symmetry, i.e., $\mathcal{M}=i\kappa_{x}%
\sigma_{y}\tau_{y}$ with mirror plane along the phase domain wall. obviously,
in-plane $y$-direction magnetic field can break the mirror symmetry and split
the pair of MZMs. We suggest further experiment to verify this prediction.
Another issue is the temperature effect. The zero-energy state can only be
observed below a temperature which is quite lower than $T_{c}$. This behavior
can also be understood by our self-consistent calculations, from which, one
can find that the QPT is quenched at quite lower temperature than $T_{c}$ (See
SMs\cite{SM} for details). The third interesting property is the behavior of
the tunneling spectra under different tunneling transmission by tuning
tunneling barrier. The experiment of trilayer heterostructure observed a
saturated value of 0.22 in unit of $2e^{2}/h$ which is much smaller than
vortex case with nearly a saturated value of 1 in unit of $2e^{2}/h$. The
smaller value 0.22 strongly indicates the observed zero-energy state in the
trilayer heterostructure experiment comes from the topmost sublayer of 1-UC
Bi$_{2}$Te$_{3}$ as we propose, and the MnTe layer inevitably lower the
tunneling transmission. Namely, the MZMs are stealth as we have pointed out.

Based on uncovering and understanding the underlying physics for the
zero-energy states, we improve and propose new heterostructure to realize the
MZMs, as shown in Fig. \ref{fig-mzm} (e). In the left plot of Fig.
\ref{fig-mzm} (e), the MnTe layer is not necessary, and substrate Fe(Te,Se)
can also be replaced by other superconductor such as NbSe$_{2}$. The only
matter is to introduce a FM impurity or defect on the topmost layer of
Bi$_{2}$Te$_{3}$ thin film. The superiority of heterostructure in Fig.
\ref{fig-mzm} (e) is the coupling between FM impurity or defect and top
surface of Bi$_{2}$Te$_{3}$ thin film can be finely tuned. It would provide a
method to manipulate the MZMs. The heterostructure can further be simplified
by adopting a single superconducting topolocial material to replace the
Bi$_{2}$Te$_{3}$/ superconductor heterostructure, as shown in the right plot
of Fig. \ref{fig-mzm} (e).

Different scenarios can be argued to generate zero-energy state. The first one
is the spontaneous vortex state. In this case, The MZMs can be taken as end
states of one-dimensional vortex line. When the line is very short like the
thin film case, the two end states can couple and split\cite{splitting}. In
our case, the pair of MZMs lies in the phase domain wall around the defect in
the topmost sublayer of 1-UC Bi$_{2}$Te$_{3}$. Thus, there is no such
coupling, and they are robust. The second one is the possible magnetic
skyrmion. In such a case, the theory has proven only the skyrmion with
topological charge $Q=2$ can bind a MZM\cite{sky-MZM-1,sky-MZM-2,sky-MZM-3}.
Usually, $Q=2$ skyrmion is not stable and split into two $Q=1$ skyrmions.
Besides, the DFT calculations give the AFM-type not FM-type exchange coupling,
and the latter is necessary to form the familiar skyrmions. Furthermore, the
size, shape and density of the skyrmions are fragile to the change of external
magnetic field\cite{sky-split-1,sky-split-2,sky-split-3,sky-split-4}. Thus,
the skyrmion scenario is not preferable.

In conclusion, we reveal the underlying physics of the robustness of
zero-energy state in a trilyer heterostructure MnTe/Bi$_{2}$Te$_{3}%
$/Fe(Te,Se), and find that there are two types of quasi-particles. The first
one is the YSR state from the FM Mn-Bi antisite defect itself and the second
one is MZM from the superconducting phase domain induced by the defect. We
show that consideration of both types of quasi-partcles can explain all the
experimental observations. Furthermore, we propose some more simple platforms
with superiority to generate and finely tune MZMs. Our studies provide new
strategies to explore Majorana physics.

\begin{acknowledgments}
The authors thank J. P. Hu, Z. Fang, C. Fang, P. Zhang, X. X. Wu, S. B. Zhang,
S. S. Qin, F. W. Zheng, H. F. Du, L. Shan, Z. Y. Wang, S. C. Yan and X. Y. Hou
for helpful discussions. This work was financially supported by the National
Key R\&D Program of China (Grants No. 2022YFA1403200, No. 2017YFA0303201),
National Natural Science Foundation of China (Grants No. 92265104, No.
12022413, No. 11674331), the \textquotedblleft Strategic Priority Research
Program (B)\textquotedblright\ of the Chinese Academy of Sciences, Grant No.
XDB33030100, the Major Basic Program of Natural Science Foundation of Shandong
Province (Grant No. ZR2021ZD01). A portion of this work was supported by the
High Magnetic Field Laboratory of Anhui Province, China.
\end{acknowledgments}

\begin{widetext}
\section{First-principles calculations}

\subsection{Calculating methods}

First-principles calculations were performed by density functional theory
(DFT) using the Vienna ab initio simulation package (VASP) \cite{vasp1,vasp2}.
The plane-wave basis with an energy cutoff of 360 eV was adopted. The
electron-ion interactions were modeled by the projector augmented wave
potential (PAW) \cite{PAW} and the exchange-correlation functional was
approximated by the Perdew-Burke-Ernzerhof-type (PBE) generalized gradient
approximation (GGA) \cite{PBE}. The strong correlated interaction of d-orbital
electrons in Mn is taken into account by GGA+U method, where the Hubbard U
value is adopted as 5 eV. And the spin-orbital coupling (SOC) is taken into
account during all the self-consistent calculations.

\subsection{Electronic and magnetic structure of the trilayer heterostructure}

As detected by the experiment, the 1UC MnTe shares the same crystal constant
with 1QL Bi$_{2}$Te$_{3}$. Here we treat 1QL Bi$_{2}$Te$_{3}$ as a substrate
and let the MnTe layer relax freely until the force and energy criterion is
satisfied. The final structure is shown in Fig. \ref{band}(b), the Te-Mn-Te
sandwich structure is as same as it in $\alpha$-MnTe crystal. In bulk $\alpha
$-MnTe, the magnetic order is in-plane intra-layer FM and inter-layer AFM,
which is usually called A-AFM. However when it is reduced to a single layer,
the ground magnetic state tends to be in-plane AFM. As shown in Table.
\ref{MAE}, \begin{table}[h]
\caption{Relative energy of different magnetic structure. The unit is meV/Mn.}%
\label{MAE}
\setlength{\tabcolsep}{10mm}{ }
\par%
\begin{tabular}
[c]{lcr}\hline\hline
x-AFM1 & x-AFM2 & x-FM\\\hline
0 & 0.23 & 17.3\\\hline
z-AFM1 & z-AFM2 & z-FM\\\hline
0.43 & 0.43 & 18.8\\\hline
\end{tabular}
\end{table}the relative energy of different magnetic structures are
calculated, form where it is easy to find that the collinear
anti-ferromagnetic (x-AFM1, as shown in \ref{mag-band}(b)) structure has the
lowest energy, however, the energy difference is very small between two
different AFM state or directions. Compared with some typical magnetic 2D
materials whose magnetic anisotropy energy (MAE) usually reaches $\mu$eV, the
MAE of 1UC-MnTe is quite small. Additionally, the band structures of
1UC-MnTe/Bi$_{2}$Te$_{3}$ with different magnetic orders have been calculated
and the results are displayed in Fig. \ref{mag-band}. It is obvious that with
a long-range AFM magnetic order, the band will obtain a giant Rashba
splitting, which could be easily detected by experiments, however that is
negated by experiments. \begin{figure}[hptbh]
\begin{center}
\includegraphics[width=1.0\columnwidth]{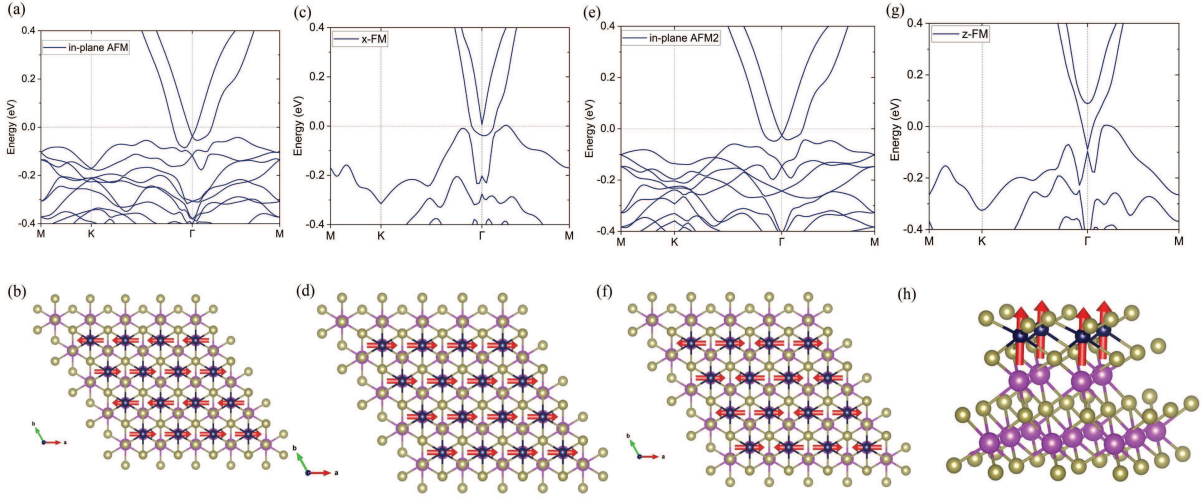}
\end{center}
\caption{Band structure of 1UC-MnTe/Bi$_{2}$Te$_{3}$ with different magnetic
structure.}%
\label{mag-band}%
\end{figure}

So here we treat the MnTe layer as spin fluctuation instead of any long-range
magnetic order. Under this setting, the band structure is obtained and shown
in Fig. \ref{band}(c) where the red dots label the weight of the surface
atomic layer's contribution. The surface state of MnTe/Bi$_{2}$Te$_{3}$ is a
typical massive Dirac cone. \begin{figure}[ptbh]
\begin{center}
\includegraphics[width=1.0\columnwidth]{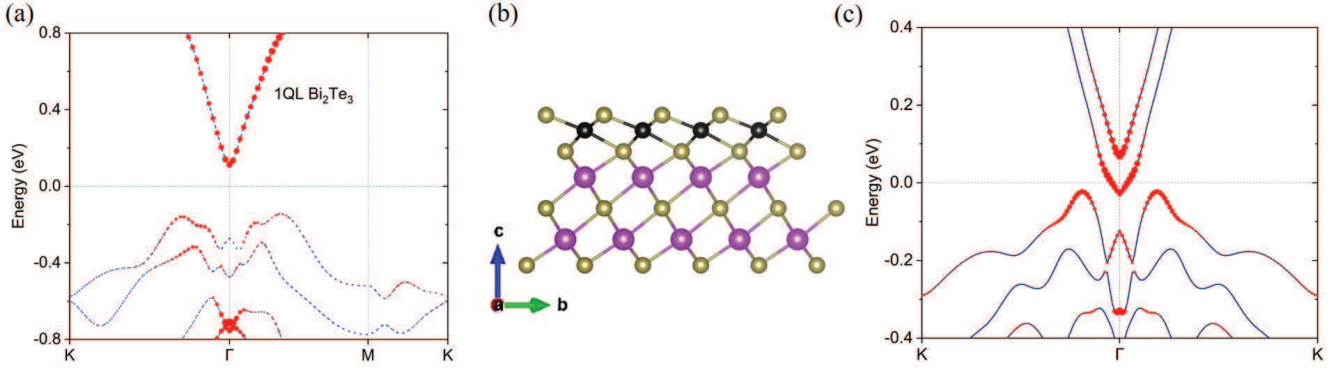}
\end{center}
\caption{Band structure of 1QL Bi$_{2}$Te$_{3}$ and 1UC-MnTe/Bi$_{2}$Te$_{3}$
without long-range magnetic order.}%
\label{band}%
\end{figure}

For 2UC-MnTe structure, due to the inter-layer interaction it tends to form a
A-AFM magnetic structure. Adopting this order, the band structure of
2UC-MnTe/Bi$_{2}$Te$_{3}$ is obtained and shown in Fig. \ref{2UC-AFMx} and
Fig. \ref{2UC-AFMz}, both of them is relatively trivial and that should the
reason why Majorana cannot be induced. Also the band structure of 2UC-MnTe
without long-range magnetic order has been considered and the results are
shown in the main text. \begin{figure}[ptbh]
\begin{center}
\includegraphics[width=1.0\columnwidth]{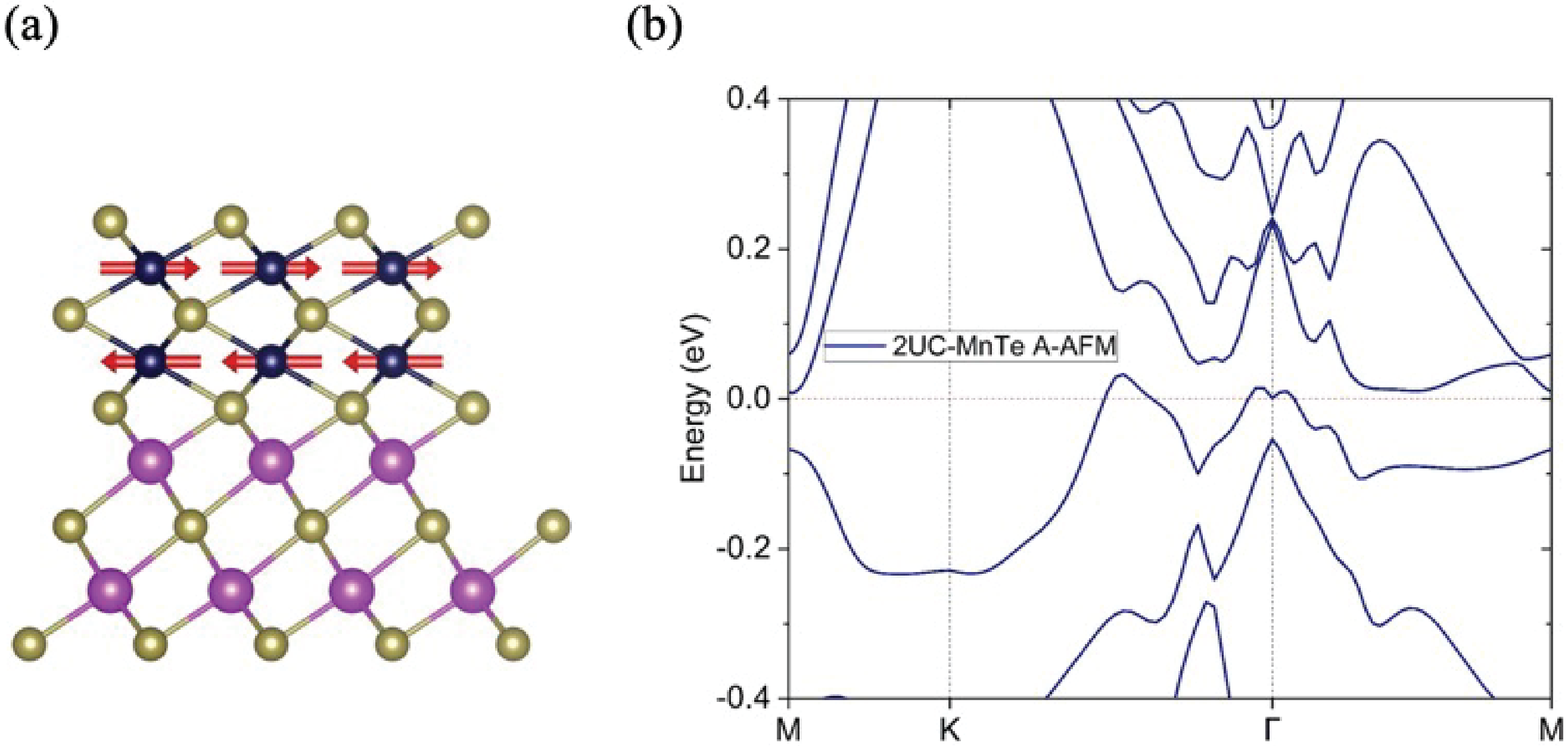}
\end{center}
\caption{Band structure of 2UC-MnTe with A-AFMx magnetic order.}%
\label{2UC-AFMx}%
\end{figure}

\begin{figure}[ptbh]
\begin{center}
\includegraphics[width=1.0\columnwidth]{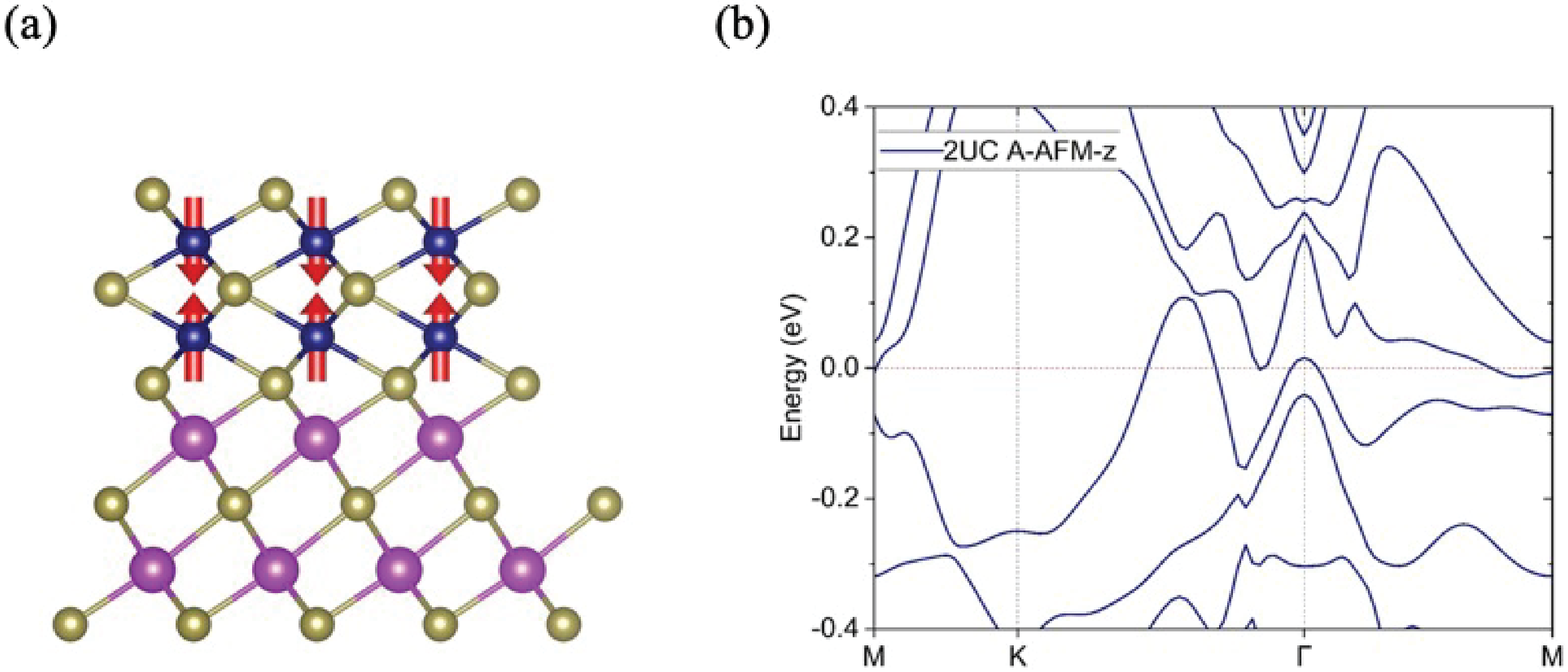}
\end{center}
\caption{and structure of 2UC-MnTe with A-AFMz magnetic order.}%
\label{2UC-AFMz}%
\end{figure}We have investigated several typical defects that could exist in
this heterostructure, which includes Mn atom vacancy, Mn-Bi antisite and Bi-Mn
substitution. The results show that the Mn-Bi antisite can easily introduce a
strong magnetic moment near the surface. The exchange interaction takes effect
between the local magnetic moment and electrons of surface states. That is
quite similar with the interstitial impurity in iron-based superconductor
Fe(Te,Se), the Mn/Fe atom is located at the center of Te atoms. The magnetic
moment of Mn impurity is obtained as 4.8 $\mu_{B}$.

\section{Theoretical Model}

\label{theo}

\subsection{The effective Hamiltonian}

At the ultra-thin limitation of a topological insulator, for example 1QL
Bi$_{2}$Te$_{3}$ \cite{massive-Dirac,massive-Dirac-2}, its surface state can
be derived as
\begin{equation}
H_{sur}=Ak^{2}+(\Lambda_{0}-Bk^{2})\sigma_{z}\kappa_{z}+v_{F}(k_{x}\sigma
_{y}-k_{y}\sigma_{x}).\label{h0}%
\end{equation}
Here, $A$ and $B$ are two constants, $\sigma$ and $\kappa$ are Pauli matrices
which spans the spin and psudo-valley space, respectively. $\Lambda_{0}$ is
the gap induced by the hybridization between surface states. To better
understand the picture of this model, we can neglect the k-quadratic term
first, and then Eq. (\ref{h0}) could be reduced to an effective Hamiltonian
as:
\begin{equation}
H_{eff}=v_{F}(k_{x}\sigma_{y}-k_{y}\sigma_{x})+\Lambda_{0}\sigma_{z}\kappa
_{z},\label{heff}%
\end{equation}
or in a matrix form as:
\begin{equation}
H_{eff}=\left(
\begin{array}
[c]{cc}%
h_{+} & 0\\
0 & h_{-}%
\end{array}
\right)  \label{hMatrix}%
\end{equation}
where $h_{\pm}=v_{F}(k_{x}\sigma_{y}-k_{y}\sigma_{x})\pm\Lambda_{0}\sigma_{z}%
$. It is obvious that both $h_{+}$ and $h_{-}$ describes a massive Dirac cone
which is gapped by an external magnetic field, however, the difference is the
direction of this magnetic field is opposite. So this hybrid interaction can
be equivalently regarded as the existence of a pseudo-magnetic field. As shown
in Fig. \ref{pMF}, we can assume that two Dirac cones are placed in two
opposite pseudo-magnetic field (pMF) and thus gain an energy gap whose
amplitude is $\pm\Lambda_{0}$. And to preserve the time reversal (TR)
symmetry, the two massive Dirac cones must locate in two opposite momentum
point, i.e. $k$ and $-k$ respectively. For example, strain induced pMF in
graphene has opposite directions when applied to the Dirac cone located at $K$
and $-K$ point \cite{pMF-1}. While the massive Dirac cones in ultra-thin
topological insulator are degenerate at $\Gamma$ point. \begin{figure}[ptbh]
\begin{center}
\includegraphics[width=1.0\columnwidth]{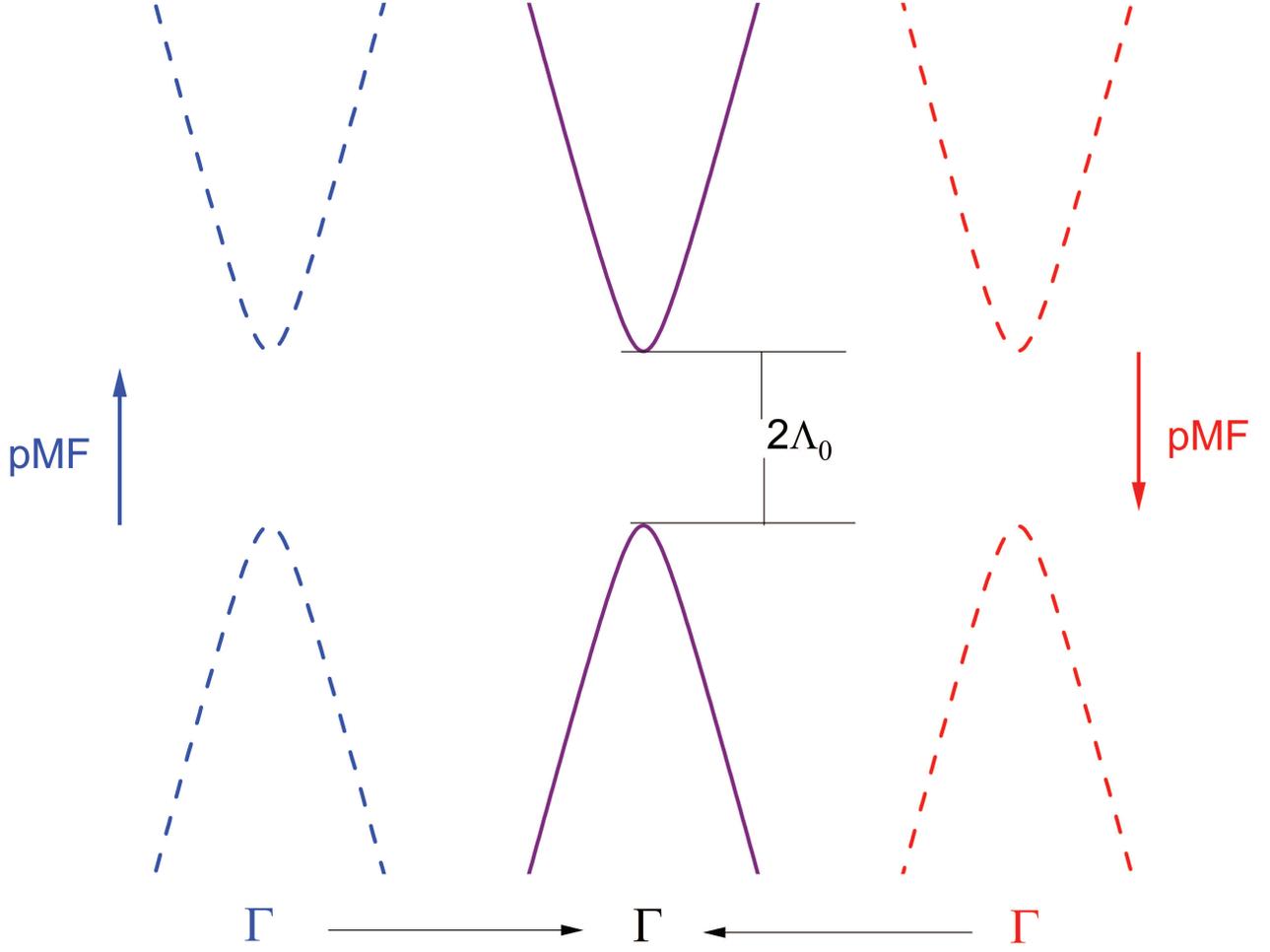}
\end{center}
\caption{Schematic diagram of pseudo-magnetic field, the blue and red dashed
one is the gapped Dirac cone belongs to the top and bottom surface,
respectively, while in k-space they are degenerated at $\Gamma$ point.}%
\label{pMF}%
\end{figure}So under the TR operation, the two massive Dirac cones will map to
each other. Therefore, the TR operator should be $\hat{T}=i\sigma_{y}%
\kappa_{x}K$ on the basis $(c_{1\uparrow},c_{1\downarrow},c_{2\uparrow
},c_{2\downarrow})$ where the subscript $1(2)$ labels two psudo-valleys and
$\uparrow(\downarrow)$ labels the spin.

At the existence of an external magnetic field, the two massive Dirac cones
correspondingly have different response, depending on the external magnetic
field is parallel or antiparallel to its pMF. For the parallel condition,
obviously, the gap is enlarged by the external field; while for the
antiparallel condition, the gap will minish to close and then reopen, this
process is actually a topological phase transition. We can estimate the
critical magnetic field of the topological phase field by this equation
\begin{equation}
\label{criticalMF}g \mu_{B} B_{c} = \Lambda_{0}%
\end{equation}
where $g$ is the Lande factor of electrons, $\mu_{B}$ is the Bohr magneton. In
1QL Bi$_{2}$Te$_{3}$, $\Lambda_{0}\simeq180 meV$ which corresponds to a
critical magnetic field $B_{c}\sim1500\ T$; In 1UC MnTe/Bi$_{2}$Te$_{3}$,
$\Lambda_{0}\simeq45 meV$ which corresponds to a critical magnetic field
$B_{c}\sim375\ T$. In above two systems, the critical magnetic field is
extremely high where the system have already entered the Landau-level range.
So when the external magnetic field is not so large, the surface state of 1QL
Bi$_{2}$Te$_{3}$ or 1UC-MnTe/Bi$_{2}$Te$_{3}$ is very stable. Of course, the
degenerate of two massive Dirac cones will be lifted by the external magnetic
field, but the splitting energy is small. Additionally, TR symmetry is broken
by the external magnetic field which results the TR partners $h_{+}$ and
$h_{-}$ decouple with each other.

\subsection{Emergence of Majorana Zero-energy Mode}

The sub-surface defect can induce a strong local magnetic moment near the
surface, which will be seen in the next section. The exchange interaction will
trigger a quantum phase transition which makes a $\pi$-phase area around the
defect when the surface enters superconductivity. At the boundary of 0 and
$\pi$-phase area, there exists the bound state whose wave function obeys the
anti-periodic condition \cite{anti-peri}
\begin{equation}
\Psi(r,\theta)=-\Psi(r,\theta+2\pi),
\end{equation}
and satisfies the BdG equation
\begin{equation}
\label{BdG}H_{BdG}(k\rightarrow-i\nabla_{r}) \Psi(r,\theta) = E \Psi
(r,\theta).
\end{equation}
Here, the BdG Hamiltonian can be expressed as
\begin{equation}
H_{BdG}=H_{BdG}^{+}\oplus H_{BdG}^{-}%
\end{equation}
with
\begin{equation}
\label{h+-}H_{BdG}^{\pm}=(h_{\pm}-\mu)\tau_{z}+\Lambda_{0}\sigma_{z}%
+\Delta_{0}\tau_{x}%
\end{equation}
the intra-surface pairing is considered and the basis is $\Psi^{\dag
}=(c_{1k\uparrow}^{\dag},c_{1k\downarrow}^{\dag},c_{1-k\downarrow
},-c_{1-k\uparrow},c_{2k\uparrow}^{\dag},c_{2k\downarrow}^{\dag}
,c_{2-k\downarrow},-c_{2-k\uparrow})$. That means we can separately solve the
BdG equation for these two independent parts. The solution of $h_{+}$ or
$h_{-}$ is equivalent to the condition where an external magnetic field
exists, which we have solved in our previous work. Here the results are
displayed
\begin{align}
u_{\uparrow}(r)  &  = c_{1}J_{-\frac{1}{2}}(k_{F}r)e^{-\frac{r}{\xi_{0}}%
}\label{solv1}\\
u_{\downarrow}(r)  &  = c_{2}J_{\frac{1}{2}}(k_{F}r)e^{-\frac{r}{\xi_{0}}},
\label{solv2}%
\end{align}
where $k_{F}=\frac{\sqrt{\mu^{2}-\Lambda_{0}^{2}}}{v_{F}}$, $\xi_{0}%
=\frac{\Delta_{0}}{v_{F}}$, $c_{1}$ and $c_{2}$ are two normalized constants.

At the presence of an out-of-plane external magnetic field, a Zeeman term is
added to the Hamiltonian
\begin{equation}
H_{Z}=g \mu_{B} \sigma_{z} B_{z} = m_{z}\sigma_{z}.
\end{equation}
Usually, $m_{z}\ll\Lambda_{0}$, thus the solution is lightly influenced.

In real materials, sometimes the quadratic term cannot be ignored. Actually it
only has impact on out-of-plane spin texture, and the in-plane helical spin
texture is not changed so that the effective p-wave superconductivity still
can be induced, which is verified by experiments. And our numerical
calculations will prove that the existence of Majorana zero-energy mode is not
disturbed by these quadratic term at all.

In our previous work, we have proven that the 0-$\pi$ disk junction is
equivalent to the vortex condition. So the inference is naturally raised that
Majorana mode also exists at the core of a vortex in this heterostructure.
However, experimental observation shows that there is no zero-energy state in
the vortex. That can be explained by the tunneling between two Majorana
fermions which locates at the top and bottom surface, respectively. In vortex
condition, the two Majorana fermions can be seen as end states of a 1D vortex
line, when the line is long enough two end states will not overlap with each
other, however when the line is short, equivalently the topological insulator
is too thin, one Majorana fermions can tunnel to another. A simple effective
model is applied to describe this process, that is
\begin{equation}
H_{t}=\varepsilon_{0}\sum_{i=1,2}\gamma_{i}^{\dag}\gamma_{i}+t(L_{z}%
)\sum_{\langle i j \rangle}\gamma_{i}^{\dag}\gamma_{j}%
\end{equation}
where $\gamma_{i,j}$ is the annihilation operator of the Majorana fermion,
$\varepsilon_{0}$ is the intrinsic energy of the Majorana fermion which is set
as 0, and $t(L_{z})$ is the hopping integral between two Majorna fermions
which is a function of the thickness of topological insulators. As derived
from Ref. \cite{splitting}, $t_{L_{z}}$ exponentially decays with the
thickness $L_{z}$ increasing. So when the thickness is large enough, the
energy spectrum of the vortex core is a robust zero-energy peak; On the
contrary, if the thickness of TI is too thin, the zero-energy peak will split
into two peaks locating at $\pm t$, respectively. That coincides well with the
experimental observation where two symmetric peaks appear in STS spectrum when
the vortex is detected.

\section{Numerical calculations}

\subsection{Quantum phase transition induced by magnetic impurity}

Now we consider a tight-binding model in a hexagonal lattice, whose
Hamiltonian is
\begin{equation}
H_{tb}=\sum\limits_{i}(\mu+\sigma m_{0})c_{i\sigma}^{\dag}c_{i\sigma}%
+\sum\limits_{\langle ij\rangle}(t_{1}-\sigma t_{2})c_{i\sigma}^{\dag
}c_{j\sigma}+i\lambda_{R}\sum\limits_{\langle ij\rangle}c_{i\sigma}^{\dag
}(\mathbf{S}^{\sigma\sigma^{\prime}}\times\vec{d}_{ij})_{z}c_{j\sigma^{\prime
}} \label{TB}%
\end{equation}
where $\sigma$ is spin index, $\langle\rangle$ denotes the nearest neighbor
site. Taken a Fourier transformation, (\ref{TB}) becomes as
\begin{equation}
H_{tb}(k_{x},k_{y})=2t_{1}h_{0}(k)+\mu+\left[  m_{0}-2t_{2}h_{0}(k)\right]
\sigma_{z}-2\lambda_{R}h_{12} \label{tb-kspace}%
\end{equation}
where $h_{11}=\cos k_{x}+2\cos\frac{k_{x}}{2}\cos\frac{\sqrt{3}k_{y}}{2}$,
$h_{12}=\sin\frac{k_{x}}{2}\cos\frac{\sqrt{3}k_{y}}{2}\sigma_{y}-\sin
\frac{\sqrt{3}k_{y}}{2}\cos\frac{k_{x}}{2}\sigma_{y}$. Expanding
(\ref{tb-kspace}) near $\Gamma$ point, we obtain the effective Hamiltonian
$h_{+}$. considering the system enters superconductivity and a magnetic
impurity located at site $i_{0}$, the total Hamiltonian can be expressed as :
\begin{equation}
H=H_{tb}+H_{d}+H_{\Delta} \label{total}%
\end{equation}
where
\begin{equation}
H_{d}=-J(c_{i_{0}\uparrow}^{\dagger}c_{i_{0}\uparrow}-c_{i_{0}\downarrow
}^{\dagger}c_{i_{0}\downarrow}),\ \ \ H_{\Delta}=\sum_{i}(\Delta c_{i\uparrow
}^{\dagger}c_{i\downarrow}^{\dagger}+h.c.) \label{self-consistent}%
\end{equation}
describes the exchange and proximity-induced superconductivity in the system.
We can perform the Bogoliubov transformation
\begin{equation}
c_{i\sigma}=\sum_{n}^{^{\prime}}(u_{i\sigma}^{n}\gamma_{n}-\sigma v_{i\sigma
}^{n\ast}\gamma_{n}^{\dagger}) \label{self-consistent}%
\end{equation}
where $^{\prime}$ denotes summation over the positive eigenvalues, and
numerically solve the equations
\begin{equation}
\sum_{j}\hat{H}_{ij}\phi_{j}=E_{n}\phi_{i}%
\end{equation}
in the Nambu spinor representation $\phi_{i}=(u_{i\uparrow}^{n},u_{i\downarrow
}^{n},v_{i\uparrow}^{n},v_{i\downarrow}^{n})^{T}$. The order parameter
$\Delta_{i}$ should be self-consistently determined as
\begin{equation}
\Delta_{i}(T)=\frac{g}{2}\sum_{n}^{^{\prime}}(u_{i\uparrow}^{n}v_{i\downarrow
}^{n\ast}+u_{i\downarrow}^{n}v_{i\uparrow}^{n\ast})\tanh(\frac{E_{n}}{k_{B}T})
\end{equation}
To get the numerical results, we have adopted a $25\times25$ lattice. The
parameters are set as $t_{1}=1$ meV, $t_{2}=0.1\ t_{1}$, $\mu=0.8\ t_{1}$,
$\Lambda_{0}=0.5\ t_{1}$, $\Delta_{0}=2\ t_{1}$, $\lambda_{R}=3.5\ t_{1}$. The
results are shown in Fig. \ref{QPT}, where the quantum phase transition
happens when $J$ crosses over $J_{c}$, meanwhile two branches of YSR states
meet at zero energy and the order parameter at the impurity mutates, as shown
in Fig. \ref{QPT}(a)(b). The consequence of QPT is the emergence of phase
domain wall around the impurity as shown in Fig. \ref{QPT}(c). As the
temperature increases, the QPT could be quenched far below the superconducting
critical temperature T$_{c}$, as shown in Fig. \ref{QPT}(d).
\begin{figure}[ptbh]
\begin{center}
\includegraphics[width=1.0\columnwidth]{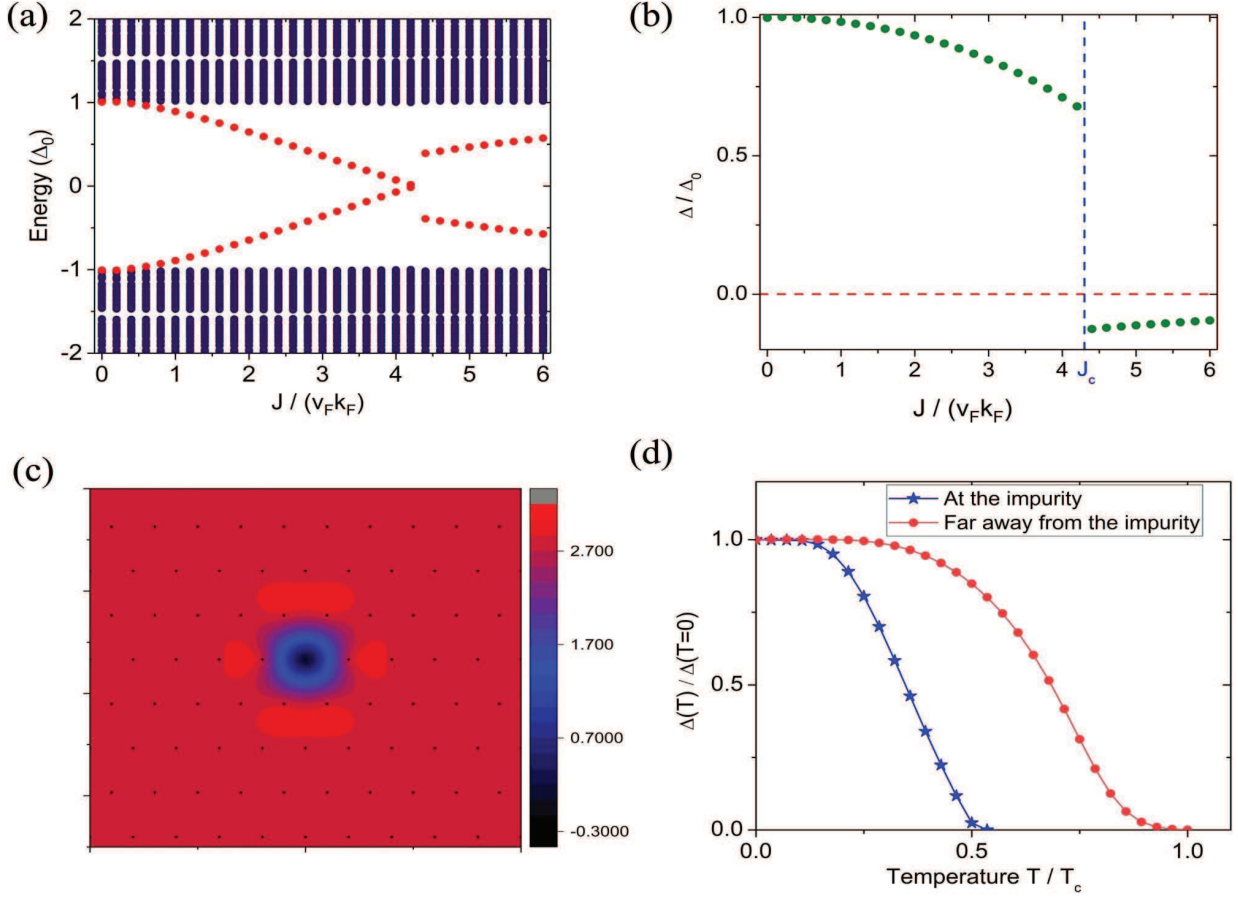}
\end{center}
\caption{(a) YSR state varies with the strength of exchange coupling between
the magnetic and substrate. (b) superconducting order parameter at the
impurity varies with $J$. (c) The spacial distribution of order parameter when
$J\gtrsim J_{c}$. And (d) order parameter at and far away from the impurity
varies with temperature, respectively.}%
\label{QPT}%
\end{figure}

\subsection{Majorana zero mode}

In Section. \ref{theo} we have proven that by adjusting the order of basis,
the BdG Hamiltonian can be block diagonal. Thus we only need two solve one
part of them. The radial form of the first-part BdG equation is
\begin{equation}
\label{A-part}H_{BdG}^{+}\psi(r)=E\psi(r).
\end{equation}
Here $H_{BdG}^{+}$ and $\psi(r)$ has the following form {\small
\begin{equation}
\label{radial}\setlength{\arraycolsep}{0.5pt} H_{BdG}^{+}= \left(
\begin{array}
[c]{cccc}%
(A-B)\nabla_{\nu,r}^{2}+\Lambda_{0}-\mu & -v_{F}(\partial_{r}+\frac{\nu+1}%
{r}) & \Delta(r) & 0\\
v_{F}(\partial_{r}-\frac{\nu}{r}) & (A+B)\nabla_{\nu+1,r}^{2}-\Lambda_{0}-\mu
& 0 & \Delta(r)\\
\Delta(r) & 0 & -(A-B)\nabla_{\nu+1,r}^{2}+\Lambda_{0}+\mu & v_{F}%
(\partial_{r}+\frac{\nu+1}{r})\\
0 & \Delta(r) & -v_{F}(\partial_{r}-\frac{\nu}{r}) & -(A+B)\nabla_{\nu,r}%
^{2}-\Lambda_{0}+\mu
\end{array}
\right)
%TCIMACRO{\U{62e2}}%
%BeginExpansion
\protect\rule{0.1in}{0.1in}
%EndExpansion%
%TCIMACRO{\U{5362}}%
%BeginExpansion
\protect\rule{0.1in}{0.1in}
%EndExpansion
\end{equation}
}
\begin{equation}
\label{wave-A}\psi(r)= \left(
\begin{array}
[c]{c}%
u_{1\uparrow}(r)\\
u_{1\downarrow}(r)\\
v_{1\downarrow}(r)\\
- v_{1\uparrow}(r)
\end{array}
\right)  .
\end{equation}
where $\nabla_{\nu,r}^{2}=\partial_{r}^{2}+\frac{1}{r}\partial_{r}-\frac
{\nu^{2}}{r^{2}}$. And applying the Majorana condition
\begin{align}
\label{c2}u_{1\uparrow}(r)  &  = -v_{1\uparrow}(r)\\
u_{1\downarrow}(r)  &  = -v_{1\downarrow}(r)\\
l  &  = 0 , \label{c3}%
\end{align}
we obtain the 1D radial differential equation list as
\begin{align}
\left[  (A-B)(\partial_{r}^{2}+\frac{1}{r}\partial_{r}-\frac{1}{4 r^{2}%
})+\Lambda_{0}-\mu\right]  u_{\uparrow}-\left[  v_{F}(\partial_{r}+\frac
{1}{2r})+\Delta_{0}\right]  u_{\downarrow}  &  = 0\\
\left[  (A+B)(\partial_{r}^{2}+\frac{1}{r}\partial_{r}-\frac{1}{4 r^{2}%
})-\Lambda_{0}-\mu\right]  u_{\downarrow}+\left[  v_{F}(\partial_{r}+\frac
{1}{2r})+\Delta_{0}\right]  u_{\uparrow}  &  = 0
\end{align}
It is hard to get the analytic solution. Here we will discuss some special
conditions where analytic solutions can be given : (a) $B=0$, it's equivalent
to a Rashba system gapped by an external magnetic field, and the requirement
of a Majorana mode yields $\mu^{2}+\Delta_{0}^{2}>\Lambda_{0}^{2}$; (b)
$A=B=0$, the solution is listed as (\ref{solv1}) and (\ref{solv2});

Next we will perform the difference method \cite{diff-method} to get the
energy spectrum and check the existence of Majorana zero-energy mode. Note
that
\begin{align}
\partial_{r}u(r)  &  = \mathop{limit}\limits_{h \rightarrow0} \frac
{u(r+h)-u(r)}{h}\\
\partial_{r}^{2}u(r)  &  = \mathop{limit}\limits_{h \rightarrow0}
\frac{u(r+2h)+u(r-2h)-u(r)}{4h^{2}} ,
\end{align}
we can divide the radius into discrete points and apply this transformation
\begin{equation}
\label{transfo}\left(
\begin{array}
[c]{c}%
u_{1\uparrow}(r)\\
u_{1\downarrow}(r)\\
v_{1\downarrow}(r)\\
- v_{1\uparrow}(r)
\end{array}
\right)  = \frac{1}{\sqrt{r}}\left(
\begin{array}
[c]{c}%
w_{1\uparrow}(r)\\
w_{1\downarrow}(r)\\
z_{1\downarrow}(r)\\
- z_{1\uparrow}(r)
\end{array}
\right)
\end{equation}
Under the new basis
\[
\psi^{\prime}=\left[  \cdots\phi(r_{i}),\phi(r_{i+1}),\cdots\right]  ^{T}%
\]
where $\phi(r_{i})=\left[  w_{1\uparrow}(r_{i}),w_{1\downarrow}(r_{i}%
),z_{2\downarrow}(r_{i}),-z_{2\uparrow}(r_{i})\right]  $ and $r_{i}=r_{0}+i
h$, Eq. \ref{A-part} can be written as
\begin{equation}
\label{discrete-BdG}\left(
\begin{array}
[c]{cccccc}%
H_{on}(r_{i}) & H_{NN} & H_{NNN} & 0 & 0 & \cdots\\
H_{NN}^{\dag} & H_{on}(r_{i+1}) & H_{NN} & H_{NNN} & 0 & \cdots\\
H^{\dag}_{NNN} & H_{NN}^{\dag} & H_{on}(r_{i+2}) & H_{NN} & H_{NNN} & \cdots\\
0 & H^{\dag}_{NNN} & H_{NN}^{\dag} & H_{on}(r_{i+3}) & H_{NN} & \cdots\\
0 & 0 & H^{\dag}_{NNN} & H_{NN}^{\dag} & H_{on}(r_{i+4}) & \cdots\\
\vdots & \vdots & \vdots & \vdots & \vdots &
\end{array}
\right)  \left(
\begin{array}
[c]{c}%
\phi(r_{i})\\
\phi(r_{i+1})\\
\phi(r_{i+2})\\
\phi(r_{i+3})\\
\phi(r_{i+4})\\
\vdots
\end{array}
\right)  = E\left(
\begin{array}
[c]{c}%
\phi(r_{i})\\
\phi(r_{i+1})\\
\phi(r_{i+2})\\
\phi(r_{i+3})\\
\phi(r_{i+4})\\
\vdots
\end{array}
\right)
\end{equation}
where
\begin{equation}
\label{hnnn}H_{NNN}=\left(
\begin{array}
[c]{cccc}%
\frac{(A-B)}{4h^{2}} & 0 & 0 & 0\\
0 & \frac{(A+B)}{4h^{2}} & 0 & 0\\
0 & 0 & -\frac{(A+B)}{4h^{2}} & 0\\
0 & 0 & 0 & -\frac{(A-B)}{4h^{2}}%
\end{array}
\right)  ,H_{NN}(r)=\left(
\begin{array}
[c]{cccc}%
0 & -\frac{v_{F}}{h} & 0 & 0\\
0 & 0 & 0 & 0\\
0 & 0 & 0 & 0\\
0 & 0 & -\frac{v_{F}}{h} & 0
\end{array}
\right)
\end{equation}
and {\small
\[
\setlength{\arraycolsep}{0.5pt} H_{on}(r)=\left(
\begin{array}
[c]{cccc}%
-(A-B)\nabla^{\prime}_{\nu,r}+\Lambda_{0}-\mu & -v_{F}(\frac{\nu+\frac{1}{2}%
}{r}-\frac{1}{h}) & \Delta(r) & 0\\
-v_{F}(\frac{\nu+\frac{1}{2}}{r}-\frac{1}{h}) & -(A+B)\nabla^{\prime}%
_{\nu+1,r}-\Lambda_{0}-\mu & 0 & \Delta(r)\\
\Delta(r) & 0 & (A-B)\nabla^{\prime}_{\nu+1,r}+\Lambda_{0}+\mu & v_{F}%
(\frac{\nu+\frac{1}{2}}{r}+\frac{1}{h})\\
0 & \Delta(r) & v_{F}(\frac{\nu+\frac{1}{2}}{r}+\frac{1}{h}) & (A+B)\nabla
^{\prime}_{\nu,r}-\Lambda_{0}+\mu
\end{array}
\right)
\]
} with $\nabla^{\prime}_{\nu,r}=\frac{\frac{1}{4}-\nu^{2}}{r^{2}}+\frac
{1}{2h^{2}}$. The energy spectrum is obtained as Fig. \ref{energy}, where the
Majorana zero-energy mode appears in the zero-angular-momentum channel, i.e.
$l=0$ ($\nu$=-1/2). \begin{figure}[ptbh]
\begin{center}
\includegraphics[width=1.0\columnwidth]{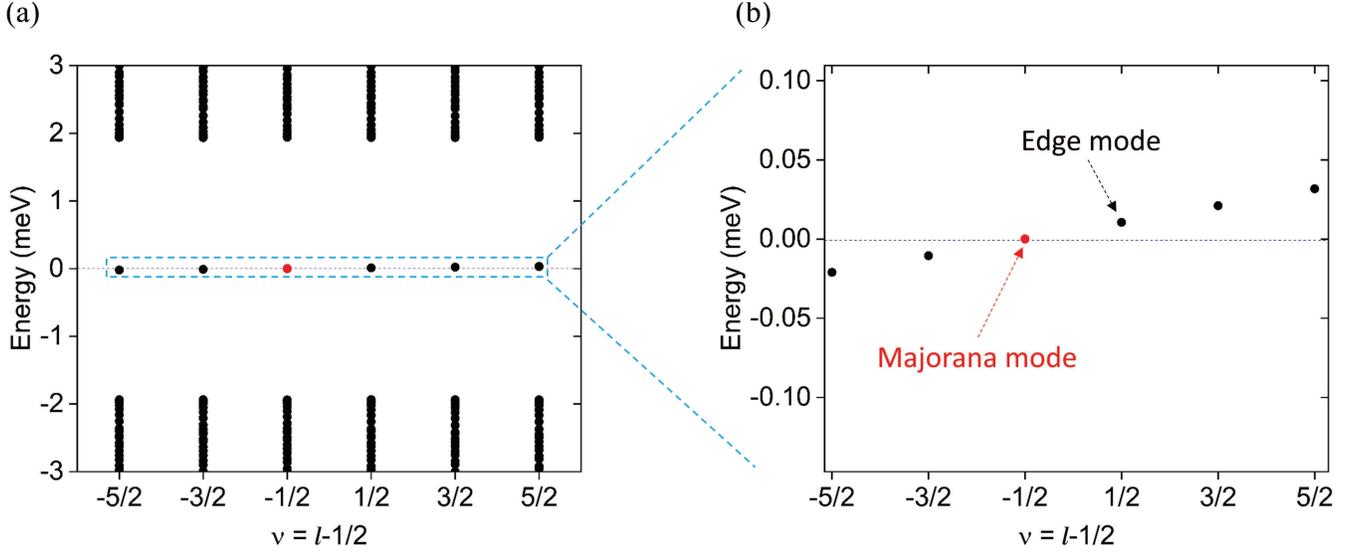}
\end{center}
\caption{(a) The energy spectrum, where $\nu$ is the pseudo angular momentum
with $\nu=l-\frac{1}{2}$. (b) is the enlarged view of the box in (a).}%
\label{energy}%
\end{figure}

Note that at $l\neq0$ ($\nu\neq-1/2$), there exist some near-zero-energy modes
as shown in Fig. \ref{energy}(b). We see they are just some trivial edge modes
originating from the finite-size-disk geometric configuration. Firstly, the
energy of these edge modes is much larger than the Majorana mode whose energy
is about 10$^{-13}$ meV in our numerical calculations; Secondly, the wave
functions of these edge modes are confined near the edge of the disk, in
reality, the radius should be infinite so these edge modes cann't emerge in
experiments. However, the wave function of the Majorana mode concentrate on
the core of the disk, i.e. around the impurity and satisfy the Majorana
condition (\ref{c1})-(\ref{c3}), as shown in Fig. \ref{wave}(a)-(d).
\begin{figure}[ptbh]
\begin{center}
\includegraphics[width=1.0\columnwidth]{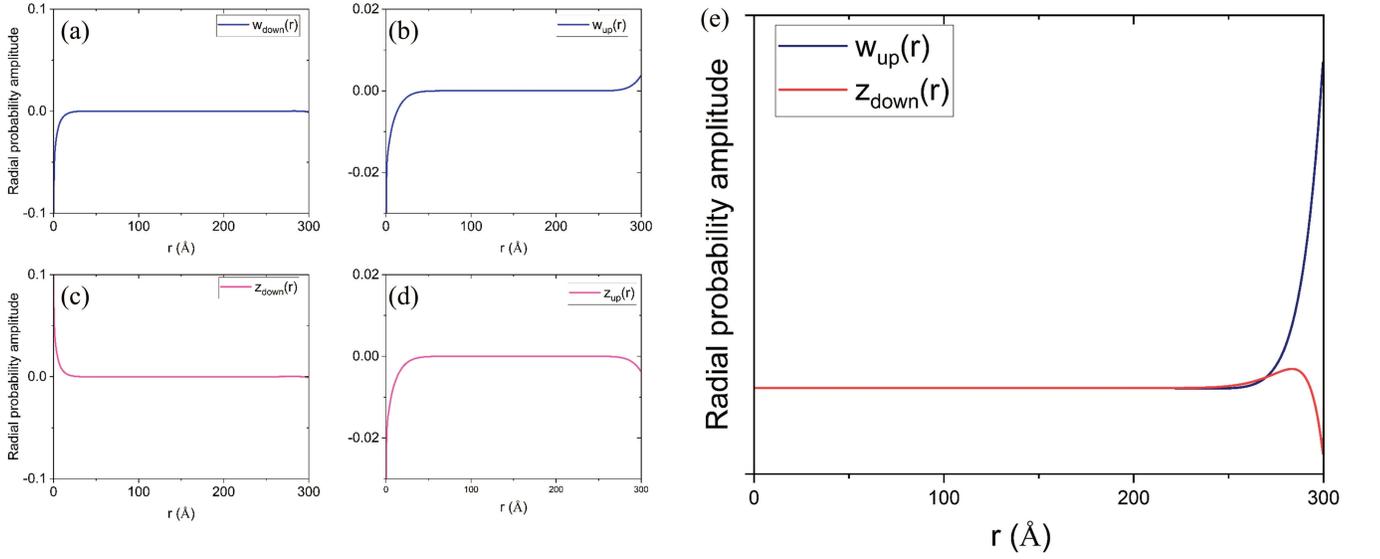}
\end{center}
\caption{Radial wave functions of the Majorana mode (a)-(d) and the trivial
edge mode (e).}%
\label{wave}%
\end{figure}

\end{widetext}

\end{document}